**Design of Atomically Precise Nanoscale Negative Differential Resistance Devices**

*Zhongcan Xiao,# Chuanxu Ma,# Jingsong Huang, Liangbo Liang, Wenchang Lu, Kunlun Hong, Bobby G. Sumpter, An-Ping Li,\* J. Bernholc\**

Z. Xiao
1) Department of Physics, North Carolina State University, Raleigh, NC 27695, USA
Prof. W. Lu, Prof. J. Bernholc
1) Department of Physics, North Carolina State University, Raleigh, NC 27695, USA
2) Computational Sciences and Engineering Division, Oak Ridge National Laboratory, Oak Ridge, TN 37831, USA
E-mail: bernholc@ncsu.edu
Dr. J. Huang, Dr. B. G. Sumpter
2) Computational Sciences and Engineering Division, Oak Ridge National Laboratory, Oak Ridge, TN 37831, USA
3) Center for Nanophase Materials Sciences, Oak Ridge National Laboratory, Oak Ridge, TN 37831, USA
Dr. C. Ma, Dr. L. Liang, Dr. A-P. Li
3) Center for Nanophase Materials Sciences, Oak Ridge National Laboratory, Oak Ridge, TN 37831, USA
E-mail: apli@ornl.gov

Down-scaling device dimensions to the nanometer range raises significant challenges to traditional device design, due to potential current leakage across nanoscale dimensions and the need to maintain reproducibility while dealing with atomic-scale components. Here we investigate negative differential resistance (NDR) devices based on atomically precise graphene nanoribbons. Our computational evaluation of the traditional double-barrier resonant tunneling diode NDR structure uncovers important issues at the atomic scale, concerning the need to minimize the tunneling current between the leads while achieving high peak current. We propose a new device structure consisting of multiple short segments that enables high current by the alignment of electronic levels across the segments while enlarging the tunneling distance between the leads. The proposed structure can be built with atomic precision using a scanning tunneling microscope (STM) tip during an intermediate stage in the synthesis of an armchair nanoribbon. An experimental evaluation of the band alignment at the interfaces and an STM image of the fabricated active part of the device are also presented. This combined theoretical-experimental approach opens a new avenue for the design of nanoscale devices with atomic precision.

## 1. Introduction

Designing band alignment to manipulate electronic transport behaviors across an interface is the key to achieving novel functionalities in semiconductor junctions and heterostructures (HSs). Esaki's discovery of negative differential resistance (NDR) in a tunneling diode six decades ago [1] continues as a treasure trove for new device design and a testbed for novel functional materials. [2-13] The recent development of graphene and graphene nanoribbons (GNRs) offers new opportunities to design nanoscale devices and to test NDR at the atomic scale. Experimentally, following the bottom-up synthesis of atomically precise GNRs,[14] HSs based on GNRs with sub-nanometer widths and various types of band alignment were successfully fabricated.[15-21] In particular, controllable polymer-to-GNR conversion reaction was demonstrated using charge injection from a scanning tunneling microscope (STM) tip.[22, 23] This advance enables the creation of atomically precise HSs and devices based on single ribbons.

Conventional Esaki devices based on doping are hard to fabricate with sufficient control.[1] Therefore, other methods were proposed to achieve NDR. In the simplest resonant tunneling diode (RTD) configuration, a quantum dot is separated from the leads by barriers, leading to confined electron level(s), see **Figure 1**. When the bias initially increases, the source Fermi level moves closer to the confined level, leading to a current increase. At a certain bias, resonant transmission is achieved, and the current reaches a maximum. As the bias increases further, the source Fermi level moves above the resonance and the current decreases. This results in an NDR region. When the bias increases further, the source Fermi level may approach another confined level and the current increases again. The change in the energy level alignments is schematically shown in Figure 1.

For an atomic scale device, the conventional RTD design needs modification to become practical. First, the small size of the segments results in strong confinement, which limits the number of tunneling levels available at moderate voltages. Second, if the segments are chosen short, direct tunneling between leads may occur and thus wash out or eliminate the NDR. Conversely, if the segments are long enough to suppress direct tunneling, electron transmission across these regions with large electronic gap results in a very small current, which may render the device impractical. Prior works have modeled potential GNR-based NDR devices that would employ doping[24-28] or gating[29] to tune the band structure, or introduce molecular quantum dots[30, 31] or defects[32] or interfaces with varying zigzag-GNR widths.[33] However, these designs require atomic precision in device fabrication, and their feasibility has not been experimentally evaluated.

In this work, we demonstrate a practical device structure, based on armchair GNRs, to deliver a strong NDR effect. The proposed GNR based HSs consist of seven-carbon-atom wide armchair GNRs (7-aGNRs) and an intermediate structure appearing in GNR synthesis, which is partially converted GNRs with one side of the polyanthrylene converted to the GNR structure while the other side remains in the polymeric structure. The GNR/intermediate HSs are confirmed to have a type-I band alignment, and thus can be employed to design NDR devices. After illustrating the key issues of NDR device design at the atomic scale, we propose an unconventional multi-part device comprised of five short segments, which leads to a pronounced NDR with a current large enough for practical use. Such a multi-part HS can be experimentally fabricated by using in situ growth from polyanthrylene precursors and STM manipulations, ensuring an atomically precise device with well-defined, reproducible characteristics.

## 2. Results

### 2.1 GNR/intermediate Heterojunction

GNRs are grown on an Au(111) substrate with DBBA molecules as precursors by following previous reports.[14, 20, 23, 44] During growth, the DBBA molecules go through debromination and polymerization steps to form polyanthrylene, which later undergoes cyclodehydrogenation step to form GNR. During these process, it is observed that after polymerization at 470 K, an intermediate structure is formed at 600 K when the benzyne groups on only one side rotate and proceed through cyclodehydrogenation.[22] **Figure 2(a)** shows an STM image of an intermediate segment consisting of two bianthrylene units at one edge side, while the other side assumes a GNR structure, as illustrated by the structural model shown in Figure 2(a) lower panel. An intraribbon GNR/intermediate/GNR HS is thus formed.

**Figure 2(b)** shows the scanning tunneling spectroscopy (STS) curves acquired across the GNR/intermediate junction. Away from the junction interface, the 7-aGNR exhibits the characteristic band gap of about 2.9 eV, with the highest occupied and lowest unoccupied molecular orbitals of the 7-aGNR ($HOMO_g$ and $LUMO_g$) at about −1.1 eV and 1.8 eV, respectively, consistent with previous reports.[45-49] The intermediate displays a smaller band gap of about 2.4 eV, with the $HOMO_i$ and $LUMO_i$ at about −0.7 eV and 1.7 eV, respectively. A type-I band alignment between the GNR and the intermediate is clearly visualized in a two-dimensional (2D) color-coded local density of states (LDOS) map shown in **Figure 2(c)**. Thus the 7-aGNR can be used as a barrier and the intermediate part as a quantum dot in a double barrier NDR device.

**2.2 Conventional double barrier device design**

We use the GNR/intermediate junctions and undoped bulk graphene as paradigmatic probes and first consider the conventional quantum dot device model. A simple GNR/intermediate/GNR double-barrier structure is depicted in **Figure 3(a)**. Two segments of the 7-aGNR, each with a length of 4 anthrylene units, ~17.0 Å, act as barriers, and they are directly connected to the bulk graphene leads. An intermediate structure of the same length, acting as a quantum dot, is sandwiched between the barriers, giving a structure labeled as a 4-4-4 HS. The transport *I-V* curve, calculated with non-equilibrium Green's function method, is shown in **Figure 3(b)**. The current increases with the bias up to 0.8 V, then it plateaus slightly before increasing again, but no obvious NDR is found. After 0.9 V the increase is more rapid because more states participate in the transmission.

**Figure 3(c)** shows the calculated position-resolved LDOS at zero bias, where the HOMO$_i$ and LUMO$_i$ from the intermediate segment, and HOMO$_g$ and LUMO$_g$ from the GNR segment can be resolved. It gives a band gap of about 1.2 eV for the intermediate and 1.6 eV for the GNR respectively, displaying a type-I band alignment, consistent with the experimental result in Figure 2. Note that the experimental band gaps for the intermediate and the GNR are both larger than those in the simulations, which can be attributed to the well-known underestimation of the band gap in DFT calculations.[50] Interface states ($E_{L1}$, $E_{R1}$ and $E_{L2}$, $E_{R2}$) between graphene leads and 7-aGNRs at 0.5 eV below and above the quasi Fermi levels ($\mu_L$ and $\mu_R$) are also marked. Due to the strong C-C bonds between the bulk graphene and the GNRs, the interface levels are broadened and decay slowly into the GNR and the graphene regions. Since both the GNR and the intermediate segments are short, the HOMO$_i$ and LUMO$_i$ of the intermediate overlap strongly with those interface states. Therefore, the 7-aGNR fails to act as a barrier and there is no clear NDR feature in the *I-V* characteristics.

To decrease the direct overlap between the interface states and the confined states in the intermediate part, we investigate the same sandwich structure with an increased length for each segment as shown in **Figure 4**. When the lengths of the barriers are increased to 34.2 Å (8 anthrylene units), the GNRs turn into true barriers for both electron and hole transport. With an 8-6-8 HS, where intermediate's length is 25.6 Å, we observe a weak NDR feature at around 0.45 V in the calculated *I-V* curve, as shown in Figure 4(a). However, the peak-to-valley current ratio (PVCR) is only 1.1, which is too small for practical use. When the lengths of the segments are further increased in an 18-16-18 HS, with GNR lengths of 76.8 Å and intermediate length of 68.3 Å, a pronounced NDR behavior indeed emerges in the 0.65 V to 0.80 V region, as can be seen from the calculated *I-V* curve for the 18-16-18 HS device shown in Figure 4(b). The calculated LDOS maps for biases of 0.65 V (current peak) and 0.8 V (current valley) are shown in Figure 4(c) and 4(d), respectively. It is very obvious from Figure 4(c)-(d) that the 7-aGNR behaves as a potential barrier and that the intermediate segment has more confined states. The interface levels ($E_{L1}$, $E_{R1}$ and $E_{L2}$, $E_{R2}$) at ±0.5 eV are similar to those in the 4-4-4 HS, but they are spatially separated much further from the confined levels in the intermediate part, in contrast to their obvious overlap in the 4-4-4 HS. When the bias is increased to 0.65 V, the HOMOs of the intermediate are aligned with the levels of the left 7-aGNR and the LUMOs are aligned with the levels of the right 7-aGNR. As a result, the current reaches its first maximum. With further increase of the bias to 0.8 V, the levels become misaligned, the current decreases and the NDR feature appears. We also note that when the levels align with each other on the left side, they fall into the band gap of the 7-aGNR at the right side. These factors lead to a significantly smaller current, only 1% compared with the current in Figure 3(b). Although the PVCR of 1.8 is relatively large, the small current makes the 18-16-18 HS configuration impractical for applications.

**2.3 Multi-segment novel device design**

To enlarge the magnitude of the current while enhancing the favorable NDR characteristics, we propose a new device design based on five short segments. The structure and electron transport properties of the new design are shown in **Figure 5**. We use two 7-aGNR barriers and three intermediate parts, sandwiched between two graphene leads [Figure 5(a)]. Instead of connecting the 7-aGNR barriers directly to the graphene leads, we insert intermediate segments between the barriers and the leads, to better align the energy levels on the opposite sides of the barriers. The use of five segments extends the active region of the device, preventing direct tunneling between the leads while allowing the barriers to be short, thereby enlarging the current. In our paradigmatic example we choose each segment to have the length of four anthrylene units, giving a 4-4-4-4-4 HS active region. In the calculated *I-V* curve in Figure 5(b), the NDR appears at a relatively small bias with a large PVCR of 3.2, which satisfies well the practical requirement for electronic circuit applications (PVCR > 3).

Similar hybrid states to those in Figure 3 are observed at both graphene/intermediate interfaces, labeled as $E_{L1}'$, $E_{R1}'$, $E_{L2}'$ and $E_{R2}'$ in Figure 5(c)-5(e). These states extend significantly into the HS and play important roles in facilitating band alignment under different biases. At zero bias, the interface levels at the graphene/intermediate interfaces are located at ±0.4 eV and are well separated in energy from the HOMO and LUMO levels at ±0.6 eV in the central intermediate part [see Figure 5(c)]. When the bias increases to 0.38 V, the HOMO$_i$ (LUMO$_i$) levels in the central intermediate part align with levels on the left (right) side intermediate segments and the interface levels $E_{R2}$ ($E_{L1}$) on the right (left) sides [see Figure 5(d)]. Therefore, the current reaches its first maximum. When the bias increases to 0.55 V, the levels become misaligned [see Figure 5(e)] and the current decreases. With the segment size similar to those in the 4-4-4 HS, the current is also of similar magnitude. The similar levels in the outer intermediate segments [marked with dashed orange ellipses in Figure 5(c)-5(e)] that are introduced by the new 5-part design are well localized in energy and bridge the gap between the leads and the central quantum dot, resolving the dilemma in the conventional design of nanoscale resonant tunneling diodes. Therefore, the 5-part structure is qualitatively better than the 3-part structure.

**3. Discussion**

In the previous section we have shown that the new device designed with multiple segments exhibits qualitatively better NDR characteristics. In this section, we examine the robustness of these NDR characteristics against changes in the multi-segment structure. It is found that the NDR feature of the 5-part structure is highly tolerant of variations in the segment widths. In **Figure 6** we compare the *I-V* curves of 5-part structures of lengths 3-4-4-4-3 and 5-4-4-4-5, as well as an asymmetric 3-4-4-3-3 structure. They all show prominent NDRs with large currents, which allow some wiggle room for the 5-part device fabrication. Pronounced NDR regions occur in the 3-4-4-4-3 HS device from 0.32 V to 0.62 V with a PVCR of 4.0 [Figure 6(a)], in the 5-4-4-4-5 HS device from 0.3 V to 0.5 V with a PVCR of 1.9 [Figure 6(b)], and in the 3-4-4-3-3 HS device from 0.37 V to 0.70 V with a large PVCR of about 6.0 [Figure 6(c)]. These results further assert the 5-part device design as a new practical paradigm for NDR devices that stands apart from the conventional quantum dot designs. We have thus succeeded in designing a practical nanoscale structure that exhibits significant NDR under ballistic current conditions, without significant tunneling at low bias.

In fact, it is feasible to "direct write" the designed interfaces with the STM tip assisted manipulations.[22, 23] The STM tip has also been used to lift up the GNRs from a conductive substrate to measure the transport behavior.[17, 49, 51] However, due to height-dependent variations of the sample-substrate coupling,[52] the band alignment in the HS can change in the lift-up configuration. To solve this problem, the HS can be lifted from both ends with two separated tips in a 4-probe STM setup.[53, 54] In this case, the interface states in the STM tip junction would be significantly different from our designed graphene/HS interfaces. As discussed above, those interfacial states are critical for the observation of the NDR. The fabrication of a real device structure requires thus coherent connections of the designed HS to graphene leads, which should also be possible as the zigzag termini of the HS can facilitate a seamless connection with graphene as demonstrated by others.[55] Clearly, experimental confirmation of the predicted NDR will require a sequence of intricate fabrication and measurement steps. While each of the steps already has a successful experimental precedent, experimental realization and verification of the predicted device is thus a significant undertaking.

The proposed 5-part structure is a novel concept in NDR device design, compared with traditional quantum dot design or a superlattice structure suggested by Tsu and Esaki.[9] The latter has been proposed for structures that can be described by the effective mass theory, which applies only to relatively wide planar segments. The above discussion takes device design to a next level and focuses on the atomic-scale structure, in which the alignment of the nearly localized levels of the segments and interface are carefully tuned, providing a distinct NDR effect while minimizing direct tunneling between the contacts.

## 4. Summary and Conclusions

Utilizing the type-I band alignment between the 7-aGNR and the polymer-GNR intermediate, we design a nanoscale NDR device with a practical peak-to-valley current ratio and peak current. The computationally-guided design uncovers novel aspects important for atomic scale devices, concerning the need to minimize direct tunneling between the leads while maintaining sufficient peak current and PVCR. Starting from the concept of a resonant tunneling diode and controlling the confinement and interfacial levels energy matching, we discover a new, broadly applicable paradigm for atomic-scale ballistic NDR devices, based on multiple narrow segments with different band gaps. Similar design concepts can be employed to fabricate atomic-scale NDR devices based on other materials, including other GNR-based devices and those based on $MoS_2$ or $WS_2$, as well as more exotic molecular structures.

## 5. Computational and Experimental Details

*Computational methods:* Density functional theory (DFT) calculations were performed with the Quantum Espresso code,[34] using ultrasoft pseudopotentials[35] with Perdew-Burke-Ernzerhof (PBE) exchange correlation potential.[36] The energy cutoff for the plane wave basis of Kohn-Sham wavefunctions was 30 Ry, and that for the charge density was 300 Ry. The atomic structures of the GNR/intermediate HSs were relaxed until forces on atoms reached a threshold of 0.002 Ry $Å^{-1}$. The calculations of the local density of states (LDOS) employed a Gaussian smearing of 0.01 eV. The transport properties were calculated with the real space[37] implementation of the self-consistent non-equilibrium Green's function method.[38, 39] The localized orbitals were optimized variationally.[40-42] Four orbitals per atom were used with a cut-off radius of 8 Bohr. The *I-V* curves were then obtained at various source-drain biases by calculating the current from the transmission spectrum using the Landauer formula.[43]

*Experimental methods:* In experiments, an Au(111) single crystal was cleaned by repeated cycles of $Ar^+$ bombardment and annealing to 740 K before growing GNRs. The 10,10'-dibromo-9,9'-bianthryl (DBBA) molecules with a purity of 98.7% were used for materials synthesis, which were degassed at 450 K overnight in a Knudsen cell. Then, the molecules were evaporated at 485 K from the cell, while the Au substrate was held at 470 K. The sample was subsequently annealed at 470 K for 30 min and 600 K for 20 min, respectively. STM/STS characterization was performed with a home-made system at 105 K under ultrahigh vacuum conditions (base pressure better than $1 \times 10^{-10}$ Torr) with a well-cleaned PtIr tip in a constant-current mode. The d$I$/d$V$ spectra were recorded using a lock-in amplifier with a sinusoidal modulation ($f$ = 731 Hz, $V_{mod}$ = 20 mV) with the feedback-loop gain off. The polarity of the applied voltage refers to the sample bias with respect to the tip.

**Acknowledgements**

A portion of this research was conducted at the Center for Nanophase Materials Sciences (CNMS), which is a DOE Office of Science User Facility. The research was funded by grants ONR N00014-16-1-3213 and N00014-16-1-3153, and DOE DE-FG02-98ER45685. The development of the RMG code was funded by NSF grant OAC-1740309. Supercomputer time was provided by NSF grant ACI-1615114 at the National Center for Supercomputing Applications (NSF OCI-0725070 and ACI-1238993).

[#] Z.X. and C.M. contributed equally to this work.

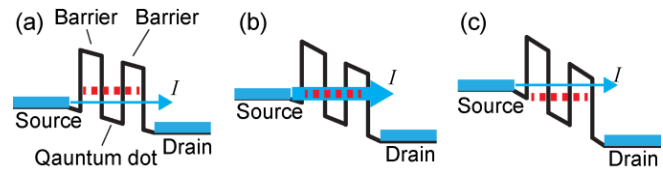

Figure. 1. Change in electron level alignment in quantum-dot-based RTD when the bias increases. (a) The first increasing current (*I*) region, (b) the maximum transmission, and (c) the decreasing current region. The dashed red line marks the energy level for resonant tunneling.

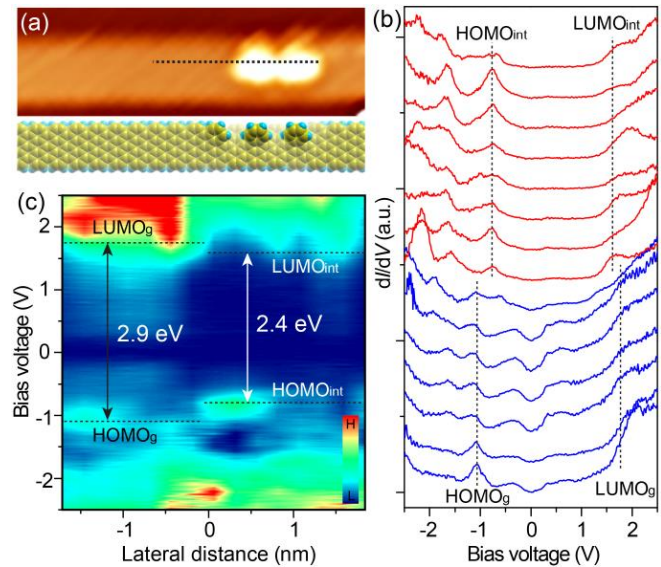

Figure 2. (a) Upper panel: STM of an HS, formed in a partially converted 7-aGNR. The partially converted segment consists of two protrusions that are intermediates between the polymer precursor and the GNR. Lower panel: Structural model of the GNR/intermediate/GNR HS. (b) The d$I$/d$V$ spectra acquired along the dotted line in the STM image of (a) (upper panel). Blue and red curves are from the 7-aGNR and the intermediate segment, respectively. The HOMO and LUMO peaks for each segment are marked. (c) Color-coded d$I$/d$V$ map plotted with measured curves in (b), giving a type-I band alignment.

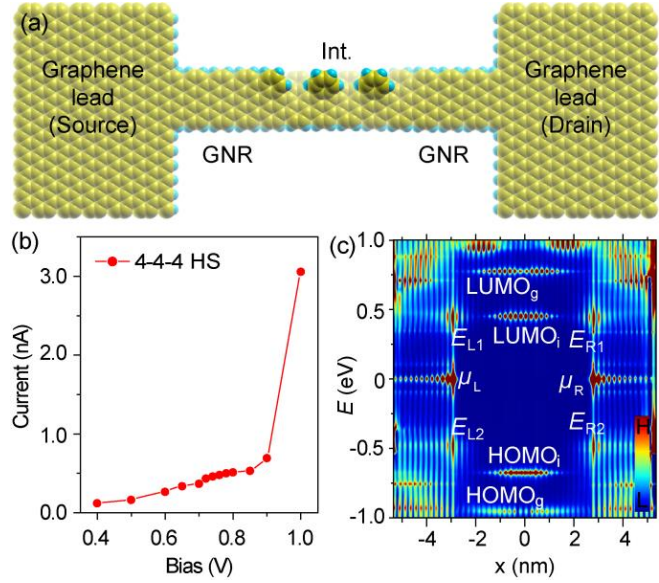

Figure 3. (a) Atomic structure of a paradigmatic quantum dot device connected to graphene leads, where an intermediate segment (labeled as Int.) serves as the tunneling channel and two 7-aGNR segments (labeled as GNR) act as barriers. All edges in the device are hydrogenated. The three-component HS is denoted as an n-m-n HS, where n, m, and n are defined by the numbers of corresponding anthrylene units in each component. (b) Calculated *I-V* curves for the 4-4-4 HS device. (c) The LDOS map along the device at zero bias. There are four interfacial levels, $E_{L1}$, $E_{R1}$ and $E_{L2}$, $E_{R2}$, at the left/right lead/GNR interfaces. HOMO$_i$ and LUMO$_i$ from the intermediate segment, and HOMO$_g$ and LUMO$_g$ from the GNR segment are also labeled. The quasi Fermi levels of the left and right leads are marked as $\mu_L$ and $\mu_R$.

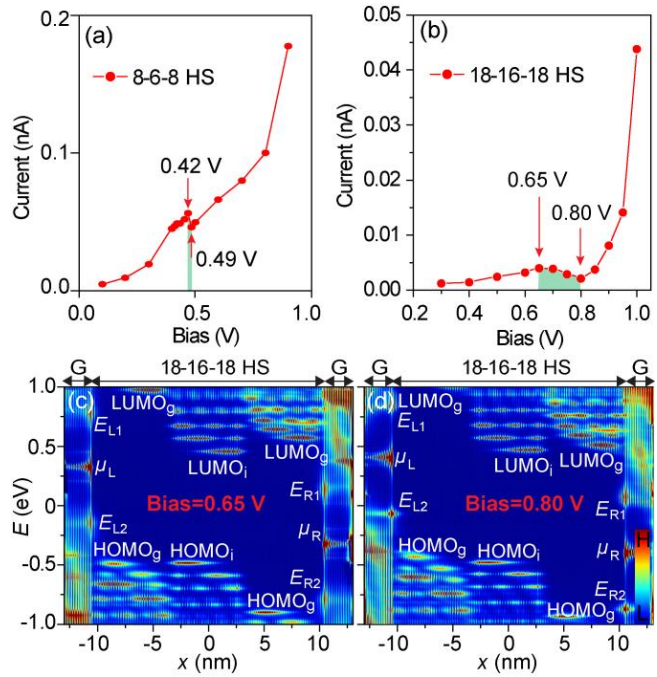

Figure 4. (a) Calculated *I-V* curves for the 8-6-8 HS device, and (b), the 18-16-18 HS device. (c) LDOS maps along the device at 0.65 V bias, and (d) at 0.80 V bias of the 18-16-18 HS device.

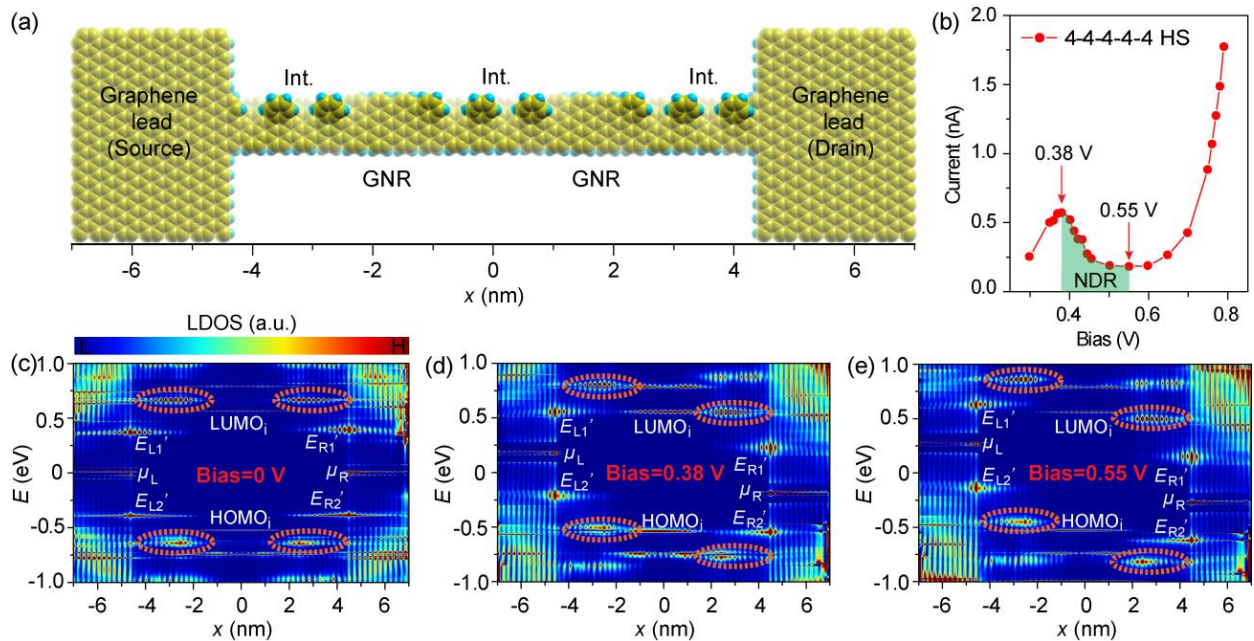

Figure. 5. (a) Atomic structure of a paradigmatic 5-part GNR/intermediate 4-4-4-4-4 HS device, which consists of three intermediate segments separated by two GNR segments, with each segment having the length of four anthrylene units, and graphene leads. (b) Calculated *I-V* curve, exhibiting NDR from 0.38 V to 0.55 V, marked with green shading. (c) LDOS maps along the device at zero bias, (d) at 0.38 V bias, and (e) at 0.55 V bias. Note that the energy levels at −0.6 eV and 0.6 eV are aligned at the bias of 0.38 V and misaligned at 0.55 V. Dashed orange ellipses mark the states emerging between the GNR and outer intermediate segments.

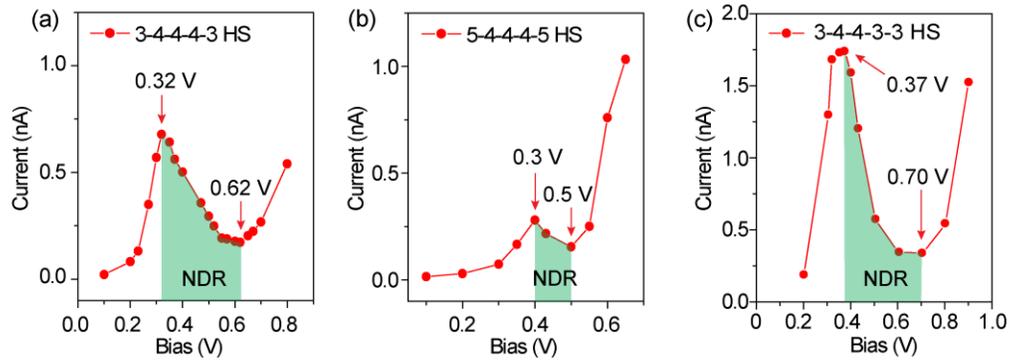

Figure. 6. (a) Calculated *I-V* curves from the 3-4-4-4-3 HS device, (b) the 5-4-4-4-5 HS device, and (c) the 3-4-4-3-3 HS device.